\documentclass{sig-alternate} 

\usepackage{graphicx}
\usepackage{amsthm}
\usepackage{amsfonts}
\usepackage{amssymb}
\usepackage{isolatin1}
\usepackage{ulem}
\usepackage{epic}
\usepackage{eepicemu}
\usepackage{subfigure}
\usepackage{stmaryrd}
\usepackage[mathscr]{euscript}
\usepackage[all]{xy}  

\newcommand{\editor}[1]{\begin{sc}#1\end{sc}}

\newtheorem{thm}{Theorem}[section]

\newtheorem{prop}{Proposition}[section]
\newtheorem{claim}{Claim}[section]
\newtheorem{cor}{Corollary}[section]
\newtheorem{fact}{Fact}[section]

\theoremstyle{definition}
\newtheorem{defn}{Definition}

\newcommand{\reffig}[1]{Figure~\ref{#1}}
\newcommand{\refsec}[1]{Section~\ref{#1}}

\newcommand{\refthm}[1]{Theorem~\ref{#1}}

\newcommand{\reffact}[1]{Fact~\ref{#1}}

\newcommand{\refprop}[1]{Proposition~\ref{#1}}

\newcommand{\refdefn}[1]{Defn.\ \ref{#1}}

\newcommand{\ignore}[1]{}
\DeclareSymbolFont{AMSb}{U}{msb}{m}{n}
\DeclareMathSymbol{\N}{\mathbin}{AMSb}{"4E}
\DeclareMathSymbol{\Z}{\mathbin}{AMSb}{"5A}
\DeclareMathSymbol{\R}{\mathbin}{AMSb}{"52}
\DeclareMathSymbol{\Q}{\mathbin}{AMSb}{"51}
\DeclareMathSymbol{\I}{\mathbin}{AMSb}{"49}
\DeclareMathSymbol{\C}{\mathbin}{AMSb}{"43}

\newcommand{\tutbox}[1]{ \begin{figure} \newlength{\pagewidth} \setlength{\pagewidth}{\columnwidth} \addtolength{\pagewidth}{-2\fboxsep} \addtolength{\pagewidth}{-2\fboxrule} \addtolength{\pagewidth}{-1em} \fbox{\begin{minipage}{\pagewidth} #1 \end{minipage} } \end{figure} }

\newcommand{\paragraphlabel}[1]{}

\newcommand{\partialrightarrow}{\rightharpoonup}

\newcommand{\indexclosure}[1]{\mathrm{Index}(#1)}
\newcommand{\semantics}[1]{\llbracket #1 \rrbracket}

\newcommand{\future}[1]{}

\newcommand{\iflong}[1]{}

\newcommand{\ifverbose}[1]{#1}

\begin{document}
%

\title{Tradeoffs in Metaprogramming
\titlenote{2006 ACM SIGPLAN Workshop on Partial Evaluation and Semantics-Based
Program Manipulation, Charleston, South Carolina, January 8-10 2006.}}
\subtitle{[Extended Abstract]}
%
%

\numberofauthors{1}
%

\author{
%
\alignauthor Todd L. Veldhuizen \\
       \affaddr{Open Systems Laboratory}\\
       \affaddr{Indiana University Bloomington}\\
       \affaddr{Bloomington, IN, USA}\\
       \email{tveldhui@acm.org}
}
\maketitle
\begin{abstract}
The design of metaprogramming languages requires appreciation
of the tradeoffs that exist between important language characteristics
such as safety properties, expressive power, and succinctness.
Unfortunately, such tradeoffs are little understood, a situation we
try to correct by embarking on 
a study of metaprogramming language tradeoffs using tools from computability
theory.  Safety properties of metaprograms are in general undecidable;
for example, the property that a metaprogram always halts and
produces a type-correct instance is $\Pi^0_2$-complete.  Although such
safety properties are undecidable, they may sometimes be \emph{captured}
by a restricted language, a notion we adapt from complexity theory.
We give some sufficient conditions and negative results on when
languages capturing properties can exist: there can be no languages
capturing total correctness for metaprograms, 
and no `functional' safety properties above $\Sigma^0_3$
can be captured.  We prove that translating a metaprogram from a
general-purpose to a restricted metaprogramming language capturing
a property is tantamount to \emph{proving} that property for the
metaprogram.  \ifverbose{Surprisingly, when one shifts perspective from
programming to metaprogramming, the corresponding safety questions
do not become substantially harder --- there is no `jump' of Turing
degree for typical safety properties.}

\end{abstract}


\terms{metaprogramming, metalanguages, program generators}



\ignore{***
\section*{Todo}

Spell check



Restore equation numbers wherever useful
***}

\section{Introduction and Overview}

\ignore{***\begin{quote}
``{\bf Generator.}  A computer routine which generates machine
instructions from skeletal definitions and from parameters supplied
by a programmer.'' --- A 1963 glossary of computer terms \cite{Fritz:CACM:1963}.
\end{quote}
***}
If one starts the clock at Konrad Zuse's insight 
that a computer could prepare its own instructions, 
metaprogramming is nearing 65 years old \cite{Ceruzzi:1998}.
Happily, it shows no sign of retiring and instead seems to grow
in prominence with each passing decade.  As a sort
of uninvited \emph{Festschrift} contribution I propose to turn a 
critical eye to it, 
investigating its nature both good and bad by characterizing
tradeoffs between facets of interest: safety, power, succinctness,
and so forth.
This paper was motivated in part by ongoing
controversy in the program generation community on how metaprogramming
tools should approach the tradeoff between safety and power, or even
if such a tradeoff exists.  Representative of such tradeoffs are the
strong safety guarantees of
MetaML \cite{Sheard:unpub:1994,Taha:TCS:2000}, the unrestrained power
(but compromised safety properties) 
of C++ template metaprogramming \cite{Czarnecki:2000},
and the moderate approach of SafeGen
\cite{Huang:GPCE:2005}.
Similar controversy exists in the design of
programming languages, where there is endless contention regarding the
`proper'
tradeoff between safety properties and expressive power.
Early programming systems of the 1950s often incorporated facilities for
syntax extensions and customized code generators (e.g., \cite{Siegal:ACM:1959}).
The 1971 programming language EL1, possibly the very first to implement
generics, 
allowed arbitrary expressions to appear where a type name was expected.
The compiler dealt with such expressions by
invoking a built-in interpreter to perform partial evaluation 
\cite{Wegbreit:SIGPLAN:1971,Brosgol:SIGPLAN:1971,Holloway:SIGPLAN:1971}.
C++ is very much in the vein of such languages: powerful generics facilities
with weak safety properties.  Type-safe languages such as ML and Java
represent an opposite philosophy, providing restricted forms of generics 
with strong safety guarantees.

A more thorough understanding of such tradeoffs would be
beneficial.
In computational complexity there is a well-established tradition of
using theory to characterize tradeoffs: between space and time, time 
and randomness,
communication and space, space and reversibility, and so forth.
A similarly methodical investigation would be useful for the field
of metaprogramming---
to map out the lay of the land, so as to have a solid theory of costs
and benefits when making design decisions.
Part of this we can borrow from the
existing literature on tradeoffs in programming languages
between power and succinctness.
For tradeoffs concerning safety properties we turn to computability
theory, which has striking explanatory power for metaprogramming tradeoffs.


It is fruitful to approach the study of meta\-programming
as the study of \emph{generalization in software}.  We view
a metaprogram as a generalization of a set of concrete instances:
a parser generator generalizes a class of parsers, for example.
To reason about many forms of generalization in a common
framework, we will adopt some uniform terminology:
\begin{itemize}
\item A \emph{metalanguage} is a language in which we define generalizations.
\item A \emph{generator} is the expression of a generalization in a metalanguage.
\item An \emph{instance} is the output produced by a generator when evaluated
on some input (e.g., parameters or source program).
\end{itemize}
The most powerful metalanguage possible is, of course, a general-purpose,
Turing-complete programming language, as used in general source-to-source
metaprogramming.
As we move to more restricted metalanguages, the
properties we can guarantee increase, and expressive power decreases.
For instance, parametric polymorphism can be usefully viewed as
defining a metalanguage for expressing generic functions:
the familiar function $\mathsf{map}$---
\begin{align*}
\mathsf{map}~:~ (\alpha \rightarrow \beta, \mathsf{list}~\alpha) \rightarrow \mathsf{list}~\beta
\end{align*}
generalizes over the concrete set of $\mathsf{map}$ functions obtained
by substituting any types for $\alpha,\beta$.  Parametric polymorphism
is highly restrictive, but if correctly implemented, type-safe.
In such examples we apply the term `generator' in a formal
sense, representing a translation from a metalanguage to concrete instances,
without the requirement that code duplication take place in the
implementation.  This terminology is
incidentally consistent with early papers on generics
that called them \emph{type generators} \cite{Liskov:SIGPLAN:1974}.

If one metalanguage can express more generalizations than another,
we say it has greater expressive power.
By considering the properties of metalanguages of varying power,
we can gain insight into the tradeoffs that exist between facets
such as power, safety, and succinctness.  

Some important themes are nicely illustrated by the
simple example of \emph{regular expressions}, which should be
familiar to most readers.
Regexps have the nice property that they \emph{capture} exactly
the computations realizable by deterministic finite automata;
this guarantees every regexp can be implemented efficiently
and we can check equivalence of their DFA representations
in $O(n \log n)$ time \cite{Hopcroft:TMC:1971}.
Practical implementations of regexps differ from
theory by introducing generalizations of frequently used patterns.
For instance one can often write
``[a-z]'' to mean ``[abcdefghijklmnopqrstuvwxyz]''.
The generalization ``[$c_1$-$c_2$]'' is not
definable in the language of regular expressions,
i.e., we cannot equate ``[$c_1$-$c_2$]'' with some regular
expression ``$s$'' that produces the appropriate sequence of characters
given arbitrary $c_1$ and $c_2$.
Instead it acts as a sort of limited escape into
a slightly stronger formalism.  This escape serves to compress
typical expressions by factoring out a
commonly occurring motif that is `incompressible' in the weaker
formalism of regular
expressions.  We can view ``[a-z]'' as
an invokation of a generator that produces the
\emph{instance} ``[abc...z]''.
The generator has the desirable property that
every instance it produces is a valid regular expression,
so the escape into the stronger formalism does not
contaminate the `nice' properties of regexps.
Against all this praise we note that when
one nears the boundary of what is `natural'
to express in regexps, monsters are quickly encountered.  For
instance, the expression---
\begin{align*}
((a|ba(aa)^\ast b)(b(aa)^\ast b)^\ast (a|ba(aa)^\ast b)|b(aa)^\ast b)^\ast \cdots \\
    \cdots (a|ba(aa)^\ast b)(b(aa)^\ast b)^\ast
\end{align*}
recognizes all strings with an even number of b's and an odd number of a's.
This set of strings is more easily recognized by
a simple program in a general-purpose language.  However, in such a setting
we have no guarantee that the language being recognized is, in fact,
regular.  Moreover, suppose we start with a program in a general-purpose 
language that recognizes a regular language, and rework it into a regular 
expression.  This reworking is tantamount to \emph{proving} 
that the set of strings accepted by the program is regular, and is as 
difficult to carry out as a proof.  The resulting loss in succinctness (i.e.,
explosion in size) cannot be bounded by any computable
function;  one can say with confidence that there are C programs
deciding regular languages that require $10^{100}$ times more
characters to write down when turned into regular expressions.
This result is due to pioneering work of Manuel Blum on succinctness
tradeoffs \cite{Blum:IC:1967}.

The themes encountered in the above example have analogues
in the design of metalanguages.  In designing a metalanguage, one
may have in mind a safety property it is desirable to guarantee:
type-safety or termination, for example.
In the ideal case a metalanguage can be found
that `captures' exactly the property: any generator
expressed in the metalanguage has the property, and any
generator compatible with the property can be expressed in
the metalanguage.  Such restricted languages generally
entail a loss in succinctness compared to a
general-purpose language.  When it is not possible to capture
exactly a safety property with a metalanguage, one is faced with 
two possible strategies.  The first is to devise a metalanguage
that guarantees the property but sacrifices expressive power;
the second is to sacrifice the safety property in favour of expressive
power.  These strategies are exemplified by the functional language
approach to generics, in which type safety is paramount, and the
approach of languages such as C++ and EL1, where one has unlimited
expressive power for generics but type safety is compromised.  In 
either strategy there is the possibility of `chasing' the property by extending
the language so as to gradually recoup expressive power or
safety.  Parametric polymorphism gives way to F-bounded polymorphism
gives way to type classes, and so forth.  In C++ there is a growing
effort to introduce some stronger (but optional) type 
safety mechanisms for generics.  
As the parade of language features extends to infinity
we may approach arbitrarily closely having both the safety property and
unlimited expressive power.

\subsection{Contributions}

This paper makes three kinds of contributions.  First, we
initiate a new research programme of characterizing tradeoffs
in metalanguages.  Second, we gather scattered information
about tradeoffs relevant to metaprogramming and make it accessible.
Third, and most importantly, we prove an array of results on
tradeoffs in metaprogramming, most previously unknown.  We prove that 
multi-stage generators are no more powerful than single-stage generators 
(\refprop{prop:multistage}).  In general, deciding safety properties
of generators is not possible.  For example, the `partial correctness'
property of whether generators written
in a general-purpose language always produce instances satisfying
a nontrivial safety property is undecidable.
The stronger property of \emph{total correctness}, i.e.,
generators always halt and produce a safe instance is
$\Pi^0_2$-complete (\refprop{prop:haltsafe}).  
The more interesting
problem is devising metalanguages that \emph{capture} safety
properties.  We prove that reworking a
generator from a general-purpose metalanguage to a restricted
metalanguage capturing a property is tantamount to \emph{proving}
that property (\refthm{thm:tantamount}).
We also give some sufficient conditions and negative results
on when languages capturing properties can exist; for example
there can be no languages capturing total correctness for
generators, nor languages capturing totality, and no `functional' safety 
properties above $\Sigma^0_3$ can be captured.  
We review the major results on succinctness in programming
languages, and show that bounded succinctness is only plausible for
languages capturing properties in $\Pi^0_2$.  
\ifverbose{
We ask when ``going meta'' implies a jump in the
degree of undecidability of a property; surprisingly,
interesting safety properties do not become much
harder when we go from programming to metaprogramming (\refsec{s:jump}).
}
Finally, we show the existence of two distinct strategies for
`safe' metaprogramming: one safe, approximating powerful languages
conservatively from below, and one powerful, approximating safe
languages from above.

\section{A catalogue of tradeoffs}

We propose to investigate metalanguages and the tradeoffs they represent,
restricting ourselves to tradeoffs suited to theoretical 
investigation.\footnote{
There is an entire family of important tradeoffs
not amenable to theoretical investigation:
those dealing with
humans, computers, and their tendency to confuse one another.
For human factors issues the reader is
recommended to the wonderful Cognitive Dimensions framework of
Thomas Green and his collaborators
that distills decades of research in
psychology of programming into digestible tradeoffs
\cite{Green:HCI:1989,Blackwell:CT:2001}.
}%
\ignore{***
We can sometimes characterize a class of
generalizations by exhibiting a metalanguage for it.
For example, the class of all computable generalizations is
exemplified by a universal (Turing-complete) programming language.
***}%
\ignore{***A metalanguage that captures a safety property guarantees us
\emph{safe generalization}.
A metalanguage capturing a complexity class (e.g.,
PTIME) lets us express any generalization whose instances can be produced
within in a particular resource bound--- providing what we shall call
\emph{resource-bounded generalization}.
***}%
 We can compare metalanguages
according to how they trade off important characteristics or \emph{facets}:
\begin{itemize}
\item The expressive power of the metalanguage, i.e.,
what generators we can define in it;
\item The safety properties we are guaranteed about the instances;
\item Succinctness, that is, how long the inputs or parameters
must be to produce instances of interest;
\item The time and space complexity of ``running'' the generator to
produce an instance.
\end{itemize}
\ifverbose{
There are additional facets not investigated in the present paper, but 
deserving of future research:
\begin{itemize}
\item The decidable properties of instances;
\item The class of problem domains for which we can write
generators that let us make programs shorter (i.e., compress them \cite{Veldhuizen:LCSD:2005}).
\item The difficulty of finding inputs to a
generator that will produce a particular instance, i.e.,
\emph{inversion} of a generator;
\item The effort required to devise an appropriate
generator, given an instance or class of instances over which we wish to
generalize.
\end{itemize}
}

Not surprisingly, we are not free to choose the best possible
properties among the above facets.  Rather, fixing one property
constrains our other choices.  For example, the following
two properties of a metalanguage are at once desirable
and irreconcilable:
\begin{enumerate}
\item The ability to describe any possible generator;
\item The safety property that every generator will produce its output
and stop.
\end{enumerate} 

The first property implies Turing-completeness; the second property
implies a metalanguage that is necessarily subrecursive 
(not Turing-complete).

\subsection{A tour of tradeoffs}
To familiarize ourselves with the nature of these tradeoffs, let us take
a brief tour through the above
facets.  To simplify we shall embark on a one-dimensional tour starting
with a universal (Turing-complete) metalanguage and
descending down a chain to primitive 
recursive generators, exponential time generators, polynomial time, and 
so forth down to very weak formalisms such as CFGs (context-free grammars)
and NFAs (nondeterministic finite automata).  CFGs and
NFAs may be used as generators by providing them `advice' on 
which nondeterministic branch to take at each step as input
(e.g., \cite{Charikar:STOC:2002}). 
This descent gives us at each stop a certain class of resource-bounded
generalizations we can define.  How do the other
facets behave as we descend from a universal language to increasingly restricted
metalanguages?
\begin{itemize}
\item With each restriction in the power of the metalanguage,
the set of generators we can express becomes, of course, smaller.
\item In a universal metalanguage we have no useful guarantees
about the behaviour of the generator; as we descend the properties
become stronger, e.g., in primitive-recursive we are guaranteed
termination.
\item As we restrict the language, the program length required to express 
generators (if they remain expressible) increases.  For at least some cases 
and if our steps are big enough, the increase in program length cannot
be bounded by any computable function.
\item The time complexity of running a generator in a universal language cannot
be bounded by any computable function.
As we descend generators are guaranteed to run faster and faster.
NFA-based generators can run in nearly linear time.
\ifverbose{
\item As we restrict the language more properties become decidable,
and they become easier to decide,
for example, in primitive-recursive we are guaranteed termination,
for NFAs we can decide whether two generators are equivalent.
\item Our ability to use generators to compress patterns is at its
peak in a general-purpose language; as we descend there may be
commonly occurring patterns and motifs that cannot be captured in the
resource-bounded metalanguage.  A classic example is Champernowne's
number
\begin{align*}
0.12345678910111213141516171819202122\cdots
\end{align*}
that is easily generated in a universal metalanguage, but
becomes incompressible in suitably restricted frameworks.  For example,
Lempel-Ziv compression is powerless against it since every subsequence
occurs equally often, a so-called `normal' real number \cite[\S 1.9]{Li:1997}.
Similarly, the failure of simple parametric polymorphism to capture
some useful patterns amongst types in generic
programming can be viewed in terms of such patterns being `incompressible'
in the restricted metalanguage.
\item Inverting a generator (finding the parameters that will cause it
to generate a particular instance) is undecidable at the higher levels,
and becomes easier as you descend.  For example, the problem of inverting
a CFG-based generator is simply parsing.  For suitably restricted forms of
CFGs the inversion problem is simply unification, for which very efficient
algorithms exist.
\item The problem of finding a generator that generalizes a given 
a set of instances starts hard and becomes easier
as the metalanguage is restricted.  In a universal language, the problem of
finding a generator for one instance is closely tied to calculating
Kolmogorov complexity, which is undecidable;  in restricted metalanguages 
it becomes possible, for example there are algorithms to
approximate the best context-free grammar producing a string
(e.g., \cite{Charikar:STOC:2002}).
}
\end{itemize}

\section{Preliminaries}

To characterize the nature of these tradeoffs we employ 
some tools of computability theory as can be found in the
introductory chapters of textbooks such as Cooper \cite{Cooper:2003},
and summarized here.
We adopt the modernized terms for computability theory 
suggested by Soare \cite{Soare:BSL:1996},
such as computably enumerable (c.e.) in place of recursively enumerable,
computable instead of recursive, and so forth.

From the vantage point of computability theory, a program represents
a function from inputs to outputs, a \emph{partial computable function}.
The computability notations for these
have a straightforward correspondence to notations
from programming language theory.  In the perhaps more familiar
`Scott brackets' notation, one writes $\semantics{p}^L x$ for the value
produced by a program 
$p$ in language $L$ running on an input $x$.  The corresponding
notation for partial computable functions is $\varphi_p^L(x)$, but usually
some universal machine (equivalently, programming language) is assumed and
one writes $\varphi_p(x)$.  

In computability theory it is traditional to consider every object 
as encoded by
a unique natural number.  The behaviour of a program is viewed as a partial
function $\varphi_p : \N \partialrightarrow \N$, from input (coded as a
natural) to output (coded as a natural), defined just for those
inputs on which the program halts.  Programs are likewise coded
by naturals, for example by enumerating all valid programs in a language
lexicographically and using a program's \emph{index} in this list.
(We will equate indices with programs throughout to minimize confusion.)
The $\varphi_p(x)$ notation is rather unfortunate for metaprogramming:
to represent the behaviour of the program generated by a
generator $g$ with parameters $x$ on an input $y$ one would write
$\varphi_{\varphi_{g(x)}}(y)$.  (The notation $\semantics{~\semantics{g}x~}y$
seems clearer).  For 
consistency with the computability
literature we will use the $\varphi$-notation.
The smaller inset box summarizes key notations for partial computable
functions.  Of these the most important are $\downarrow$ (halts)
and $\uparrow$ (diverges).

On occasion we shall employ the Church-Turing thesis to, when
presented something computable, assert there exists a
program $i$ so that $\varphi_i$ computes it.

\tutbox{
\begin{centering}
{\bf Notations for partial computable functions}

\begin{tabular}{ll}
$\varphi_p$ & Partial function computed by program $p$ \\
$\varphi_p(x)$ & Output of $p$ on input $x$ \\
$\varphi_p(x) \downarrow$ & $p$ halts on $x$ \\
$\varphi_p(x) \uparrow$ & $p$ diverges on $x$ \\
$\varphi_{p,k}(x) \downarrow$ & $p$ halts in $\leq k$ steps on $x$ \\
$\varphi_p = \varphi_{p'}$ & $p$ and $p'$ compute the same partial function \\
$\varphi_p(\N)$ & All possible outputs of $p$
\end{tabular}

\end{centering}
}

The following fact will shortly be useful; it asserts that the
partial computable functions are closed under composition.
\begin{fact}
\label{fact:composition}
If $\varphi_x$ and $\varphi_y$
are partial computable functions, there is a
program $p$ such that $\varphi_x \circ \varphi_y = \varphi_p$.
\end{fact}

In later sections we shall make use of the arithmetical hierarchy
classes $\Sigma^0_n$, $\Pi^0_n$, and $\Delta^0_n$.  These are
summarized in \reffig{f:arithmetical}.

\begin{figure}
\boxed{
\begin{minipage}{\linewidth}
\begin{centering}
{\bf The Arithmetical Hierarchy}

\end{centering}

The arithmetical hierarchy was introduced by Kleene as
a tool for classifying incomputable sets.
It consists of \emph{classes} of relations denoted $\Sigma^0_n$, $\Pi^0_n$, 
and $\Delta^0_n$, where $n \in \N$.  The bottom levels
of the hierarchy correspond to familiar classes of sets:
\begin{itemize}
\item $\Delta^0_1$ is the class of computable relations (equivalently,
decidable sets);
\item $\Sigma^0_1$ is the class of computably enumerable relations
(equivalently, c.e. sets or sets with an effective inductive definition);
\item $\Pi^0_1$ is the class of co-computably enumerable relations
(equivalently, co-c.e. sets or sets with an effective coinductive definition).
\end{itemize}
The remainder of the hierarchy is defined in terms of relations
definable by restricted forms of first-order formulas:
\begin{enumerate}
\item $\Sigma^0_0 = \Pi^0_0 = \Delta^0_0$ are the decidable relations;
\item A relation is $\Sigma^0_{n+1}$ if and only if it can be defined by a formula
of the form $\exists \overline{y} ~.~ R(\overline{x},\overline{y})$ with
$R \in \Pi^0_n$, or equivalently is c.e. relative to a $\Pi^0_n$ oracle;
\item A relation is $\Pi^0_{n+1}$ if and only if it can be defined by a formula
of the form $\forall \overline{y} ~.~ R(\overline{x},\overline{y})$ with
$R \in \Sigma^0_n$, or equivalently is co-c.e. relative to a $\Sigma^0_n$
oracle.
\item $\Delta^0_n = \Sigma^0_n \cap \Pi^0_n$
\end{enumerate}
The complement of a set $S \in \Sigma^0_n$ is a set $\overline{S} \in \Pi^0_n$
and vice versa.
The containment relations among the classes are illustrated by the
following Hasse diagram:
\begin{align*}
\xymatrix @=0.5cm {
& \vdots \ar@{-}[dr] & & \vdots \ar@{-}[dl] \\
& & \Delta^0_3 \ar@{-}[dl] \ar@{-}[dr] \\
& \Sigma^0_2 \ar@{-}[dr] & & \Pi^0_2 \ar@{-}[dl] \\
& & \Delta^0_2 \ar@{-}[dl] \ar@{-}[dr] \\
~~\text{c.e.} & \Sigma^0_1 \ar@{-}[dr] & & \Pi^0_1 \ar@{-}[dl] & \text{co-c.e.} \\
& ~~~~~~~~~~~~~~ & \Delta^0_1 & \text{decidable}
}
\end{align*}
That is, 
\begin{align*}
\Delta^0_1 \subseteq \Sigma^0_1 \subseteq \Delta^0_2 \subseteq \Sigma^0_2 \subseteq \Delta^0_3 \subseteq \cdots \\
\Delta^0_1 \subseteq \Pi^0_1 \subseteq \Delta^0_2 \subseteq \Pi^0_2 \subseteq \Delta^0_3 \subseteq \cdots 
\end{align*}
The superscript $0$ on the classes indicates the arithmetical hierarchy;
all of this hierarchy is enclosed in the first level of the
\emph{analytic} hierarchy which has superscript $1$ and is defined
in terms of second-order formulas.
\end{minipage}
}
\caption{\label{f:arithmetical}Summary of the arithmetical hierarchy}
\end{figure}

\section{A Universal Metalanguage}

\label{s:universal}

We wish to reason about metalanguages
and the tradeoffs they represent.
If generators were expressed in very different
languages, this would cause notational confusion.
We shall instead fix a universal language, and 
require that every metalanguage be a restricted
subset of the universal language.  We can do this without
loss of generality by noting that any metalanguage may be
embedded in a universal language by means of pasting together
an interpreter in the universal language with a generator and its
input as a string.
This provides a straightforward embedding of any language into
the universal language.
A typical example of such a construction is:
\begin{verbatim}
int main()
{
  print(Interpret_Lisp(
    "(lambda (x y) (plus x y))", 
    "(1 2)"));
}

string Interpret_Lisp(string prog, string input)
{
  ...
}
\end{verbatim}

Such translations are easy to produce; it is easy
to extract the original program from its embedded version; and 
the resulting program is longer by only a constant amount
(the size of the interpreter plus a little extra).
This is a
so-called ``two-part code'' construction (interpreter plus program)
\cite[\S 2.1.1]{Li:1997} and it preserves all the properties that
interest us.  
We formalize these claims as follows.

\begin{claim}
There is a universal language $U$ such that for any metalanguage $A$,
there is a computable function $f$ translating $A$ programs
to $U$ so as to satisfy these properties:
\begin{enumerate}
\item For every $A$-program $p$, the translated program $f(p)$ has
the same meaning, i.e., $\varphi^A_p = \varphi^U_{f(p)}$;
\item We can computably recognize the programs that have been translated 
from $A$, i.e., $f(\N)$ is also decidable;
\item We can computably reverse the mapping translation $f$;
\item The translation adds at most a constant factor to the program size.
\end{enumerate}
\end{claim}
\begin{proof}
(Sketch)
Since the programming language $A$ is assumed to be implementable
on a computer, we can find an interpreter for $A$ in the universal
language $U$; call this $\mathsf{Int}(p,x)$.  Define the translation
function to be the equivalent of $f(p) = \lambda x.\mathsf{Int}(p,x)$. 
Then (1) follows from the use of an interpreter; (2) is guaranteed by
the ability to examine the translated version and check that the
interpreter is exactly the interpreter for the language $A$;
(3) is ensured by
adopting an appropriate ``quotation'' mechanism for the embedded
program; (4) follows from the usual ``two-part code'' argument
\cite[\S 2.1.1]{Li:1997}
\end{proof}

\ifverbose{
We mention also that with a sufficiently powerful machine model 
we can often devise efficient interpreters that preserve
asymptotic time and space complexity up
to some small overhead, e.g., logarithmic.  
}

In the remainder of the paper we assume, for tidiness, that
metalanguages are all defined by decidable subsets of a
fixed universal language.

\section{Multistage generation}

We begin our investigation with some simple results concerning the
power of having multiple stages, rather than a single stage.
In multistage generation
one has multiple generators, each producing output taken as input by
the next (e.g., \cite{Glueck:PLILP:1995,Taha:TCS:2000}).
Assuming Church-Turing we consider each stage to be represented by a 
partial computable function.  We may then make use of the fact that by 
definition, partial computable functions are closed under composition.
The following result is then straightforward:
\begin{prop}
\label{prop:multistage}
For every multistage generator there 
is an equivalent one-stage generator.
\end{prop}
\begin{proof}
This is a simple consequence of the class of partial
computable functions being closed under composition.
Let $\varphi_{a}, \varphi_{b}, \ldots, \varphi_{n}$ be a $k$-stage
generator.  Then
by repeated application of \reffact{fact:composition},
there exists a program $\alpha$ such that 
$\varphi_\alpha = \varphi_{a} \circ \varphi_{b} \circ \ldots \circ \varphi_{n}$.
\end{proof}

\ifverbose{
In some situations we may want additional
input at each stage.
This does not offer any additional \emph{theoretical}
power, since we can provide all these inputs up-front
to the initial stage and thread unused inputs through
to later stages.  
}

Most suitable programming language mechanisms offer function
composition mechanisms that are succinct, i.e.,
the program length of the composed functions is only slightly
higher than sum of the lengths of the individual functions.  From
this we can infer that multistage generation is no more
succinct than one-stage generation.

\ifverbose{
Although multistage generators are no more powerful than
single-stage generators, they do have important practical use in
distinguishing (say) generation, compilation, load and run-time stages.  
We are, however, justified in the remainder of this paper to 
consider only the single-stage case.
}

\section{Tradeoffs in Safety}

\label{s:safety}
A primary concern in contemporary generator research
is safety of generators.  Some typical questions are:
\begin{itemize}
\item Will the generator always halt and produce an instance?
\item Will the instance produced by the generator be syntactically
well-formed?  typable?  semantically correct?
\item If a generated instance is \emph{not} going to be correct, can we detect
this and produce a sensible diagnostic message?
\item Can we devise restricted languages that `capture' useful
safety properties, e.g., every generator we write in the language
always produces type-safe instances?
\end{itemize}

As a prelude to tackling deeper questions,
in this section we review the well-known phenomena that most
functional properties of programs are undecidable, and explore
the implications for generators.  Readers familiar with Rice's
theorem are encouraged to skip ahead.


We can equate a safety property with the set of generators
having that property.
For example, if we are concerned
that the generator always halt, we can consider the set of
generators that halt on every input.  Some such sets lie in
a special class called
\emph{index sets} in computability theory \cite{Cooper:2003}.
\begin{defn}[Index set]
\label{defn:indexset}
An index set is any set of generators $B \subseteq \N$ with the following 
closure property: if $g \in B$ and $g'$ is some generator with the same 
behaviour as $g$, i.e., $\varphi_g = \varphi_{g'}$, then $g' \in B$ also.
\end{defn}
We can view an index set as a \emph{functional property} of
generators, i.e., a property definable only in terms of input-output
behaviour and without reference to (for example)
space and time consumption.

The safety property `halts on every input' can be represented by
an index set $\mathrm{Tot}$ containing every \emph{total}
computable function.

\begin{defn}[Total functions]
Define $\mathrm{Tot}$ be the index set containing all total computable 
functions, i.e., generators that halt and produce an instance for every input:
\begin{align}
\mathrm{Tot} &= \{ g \in \N ~|~ \forall x ~.~ \varphi_g(x) \downarrow \}
\end{align}
\end{defn}

The problem ``Given generator $g$, is $g \in \mathrm{Tot}$?'', i.e., wheth\-er a generator
always halts after some time and produces an output, is of course
undecidable, as are most problems concerning halting.
We might however wonder if simpler properties might be checkable,
but this turns out not to be the case because of Rice's theorem.

\begin{thm}[Rice \cite{Rice:TAMS:1953}]
\label{thm:Rice}
The only decidable index sets are $\emptyset$ and $\N$.
\end{thm}

\noindent
It is a quick step from Rice's theorem to proving that
partial correctness of generators --- if the generator halts,
it produces a safe instance --- is undecidable.
We model whatever safety property of instances we are interested
in (type-safety, syntactic correctness) 
by a set $\mathsf{SafeInstance}$ of instances satisfying that
property.

\begin{defn}
A safety property of instances is a set {\sf Safe\-Inst\-ance}
with $\emptyset \subset \mathsf{SafeInstance} \subset \N$
so there is at least one safe instance and one unsafe instance---
otherwise, the safety property would be trivial, i.e., always
true or always false.
\end{defn}

Let $\mathsf{SafeGenerator}$ be the set of generators that only
produce safe instances:
\begin{align*}
\mathsf{SafeGenerator} \equiv \{ g ~|~ \varphi_g(\N) \subseteq \mathsf{SafeInstance} \}
\end{align*}
(We write $\varphi_g(\N)$ for the image of $\N$ under $\varphi_g$, i.e.,
the set of all instances generated by $g$.)

\begin{prop}
\label{prop:safegenundec}
The problem ``Is $g \in \mathsf{SafeGenerator}$?'' is undecidable.
\end{prop}

\begin{proof}
This is straightforward:
we prove $\mathsf{SafeGenerator}$ is an index set not equal to 
$\emptyset$ or $\N$ and apply Rice's theorem.

Suppose $i,j$ are
generators with $\varphi_i = \varphi_j$ and $i \in $ {\sf SafeGenerator}.
Then $\varphi_i(\N) \subseteq \mathsf{SafeInstance}$ by definition.
Since $\varphi_i = \varphi_j$ we have $\varphi_j(\N) \subseteq \mathsf{SafeInstance}$
and hence $j \in \mathsf{SafeGenerator}$ also.  Therefore
$\mathsf{SafeGenerator}$ satisfies the closure property 
of \refdefn{defn:indexset} and is an index set.

From the definition of a safety property there exists a safe instance $s$
and an unsafe instance $\overline{s}$.  Consider the following
functions:
\begin{align*}
f(x) &= s \\
f'(x) &= \overline{s}
\end{align*}
Both functions are computable and therefore we can find programs $g$ and $g'$
that compute them.  Since $g \in $ {\sf SafeGenerator}
we have $\mathsf{SafeGenerator} \neq \emptyset$.  
From $g' \not\in \mathsf{SafeGenerator}$ we have
$\mathsf{SafeGenerator} \neq \N$.
By \refthm{thm:Rice}, $\mathsf{SafeGenerator}$ is undecidable.
\qed
\end{proof}

So, the general question of whether a generator always produces
safe instances is undecidable; this is a simple corollary to
Rice's theorem.
By finding where this problem lies in the arithmetical hierarchy
(\reffig{f:arithmetical}), we can obtain more precise details of
its undecidability.  For example, some properties
that are undecidable but $\Sigma^0_1$ can be approximated nicely--- 
we can write a program that will try to decide the property within some
time limit $t$, and as we let $t \rightarrow \infty$ we can get 
a positive answer if the property is true.\footnote{This is the
notion of a $\Delta^0_2$-approximating sequence \cite{Cooper:2003}.}
We operate under the
assumption that the safety of instances is decidable.  For example,
it is decidable whether the instance is syntactically correct
or typable; this reflects current practice.

\begin{prop}
\label{prop:safegen}
If $\mathsf{SafeInstance} \in \Delta^0_1$ then
$\mathsf{SafeGenerator} \in \Pi^0_1$.
\end{prop}
\begin{proof}
A generator is safe if and only if there is no input for which
it produces an unsafe instance.  We can therefore define the
safety property by:
\begin{align*}
\mathsf{SafeGenerator}(g) &\leftrightarrow \neg (\underbrace{\exists x ~.~ \varphi_g(x) \not\in \mathsf{SafeInstance}}_{\Sigma^0_1})
\end{align*}
This is the negation of a $\Sigma^0_1$ formula, and is therefore
$\Pi^0_1$ or co-computably enumerable.
\qed\end{proof}

It is worth noting that when we shift
from programming to metaprogramming, the safety problem 
becomes harder: if instance safety is $\Delta^0_1$, then
generator safety is $\Pi^0_1$ (and not $\Delta^0_1$).
\ifverbose{
We return to this theme in \refsec{s:jump},
where we ask when ``going meta'' is accompanied by 
ratcheting up a level in the arithmetical hierarchy.
}

That $\mathsf{SafeGenerator}$ is $\Pi^0_1$ implies we can approximate
the property by searching for counterexamples, and if a counterexample
exists, we will eventually find it.  Unfortunately if we do not find a
counterexample within a set amount of time we can conclude nothing 
about whether our generator is safe.

An even harder problem is deciding whether a generator will halt for
any input and also produce an output that is in $\mathsf{SafeInstance}$.
We show this problem is $\Pi^0_2$-complete, roughly speaking,
as hard as any $\Pi^0_2$ relation.\footnote{The exact
definition is: $\psi$ is $\Pi^0_k$-complete when
$\psi \in \Pi^0_k$, and any relation $\phi \in \Pi^0_k$ is
many-one reducible to $\psi$.  A set $\phi$ is many-one reducible to
$\psi$ when there is a computable function $f$ such that
$g \in \phi$ if and only if $f(g) \in \psi$.
}

\begin{prop}
\label{prop:haltsafe}
Suppose $\mathsf{SafeInstance}$ is decidable and
let $\psi(g)$ be the property ``$g$ halts on every input and
outputs a safe instance.''  Then $\psi(g)$ is $\Pi^0_2$-complete.
\end{prop}
\begin{proof}
First we show $\psi \in \Pi^0_2$.  We can define $\psi(g)$ by the formula:
\begin{align}
\label{e:haltsafe}
\psi(g) \leftrightarrow \underbrace{(\forall x ~.~ \underbrace{\varphi_g(x) \downarrow}_{\Sigma^0_1})}_{\Pi^0_2} \wedge \underbrace{\mathsf{SafeGenerator}(g)}_{\Pi^0_1}
\end{align}
We have a universal quantifier afront a $\Sigma^0_1$ relation, making it
$\Pi^0_2$; its conjunction with the $\Pi^0_1$ relation is
$\Pi^0_2$.

The index set $\mathrm{Tot}$ is known to be $\Pi^0_2$ complete \cite{Cooper:2003}; we
reduce deciding $\mathrm{Tot}$ to $\psi(g)$.
Given a query ``Is $f \in \mathrm{Tot}$?'' we can construct a 
function $f'(x)$ that evaluates $f$ on $x$, and if this halts, returns a safe 
instance $s \in \mathsf{SafeInstance}$.  Then $g \in \mathrm{Tot}$ if
and only if $\psi(g)$.  This is a many-one reduction (see footnote) and
therefore $\psi$ is $\Pi^0_2$-complete.
\qed\end{proof}

So, this safety property is strictly harder than deciding the
``partial correctness'' property $\mathsf{SafeGenerator}$.

The safety situation for generators written in a general-purpose
language is bleak: no nontrivial safety properties are decidable.
This fact has motivated the design of special-purpose languages for
generators that \emph{are} able to guarantee some safety property.
Of particular interest are languages that \emph{capture} a
safety property, which we investigate in the next section.

\subsection{Languages capturing properties}

As we have seen, interesting safety properties of generators written
in a universal language are undecidable.  We might conjecture that
to ensure a safety property we must sacrifice some classes of
computations that \emph{can} be done safely.  Perhaps surprisingly, this
is not always the case: we can sometimes sidestep
undecidability by designing restricted languages that `capture' the
property, in the sense that every restricted program has the property,
and conversely, for every unrestricted program with the property there 
is a functionally equivalent restricted program.
Consider for example the property $\psi$ given by
``$\varphi_p(x)$ halts for at most a finite number of inputs $x$.''
This property is undecidable (in fact $\Sigma^0_2$) but has
a trivial language capturing it: allow only programs of the
form
\begin{align}
f(x) &= 
\begin{cases}
c_1 & \text{when } x = x_1 \\
c_2 & \text{when } x = x_2 \\
~~\vdots    & ~~~~\vdots  \\
c_n & \text{when } x = x_n \\
\uparrow & \text{otherwise}
\end{cases}
\end{align}
for all finite $n$ and arbitrary constants $c_i,x_i$ for $i \leq n$.
This language clearly captures the undecidable property $\psi$.

We could capture the property
$\mathsf{SafeGenerator}$ --- ``every instance output by the generator
is safe'' ---
by a language in which any arbitrary generator can
be run, but output is filtered and any unsafe instances are replaced
with safe instances.  This captures the property in a theoretical
sense, but in practice we are fond of diagnostic messages and 
prefer compilation to always fail if $\neg \psi(p)$ holds.

There are several relevant veins of research in capture of properties 
by languages:
\begin{itemize}
\item Time- and space- complexity classes can be captured by 
restricted languages, an idea that goes back to the 1960s and has
a rich literature (e.g., \cite{Meyer:ACM:1967,Royer:1994,Jones:1997}).
There is a kind of `cheat' method, which involves a \emph{clocked}
programming language where every program comes with an attached 
statement such as ``run me for at most $c |x|^k$ steps,''
with $c$ and $k$ constants; in this case termination in polynomial
time is guaranteed.  Then there are languages that capture complexity
classes in a natural way.  
For example 
Neil Jones describes a family of languages 
with familiar constructs such as $\mathsf{while}$, $\mathsf{if}$
and $\mathsf{cons}$ that capture complexity classes like PTIME
by omitting selected constructs from the language \cite{Jones:TCS:1999}.
\item \emph{Descriptive} or \emph{Implicit computational complexity} studies 
restricted logics that
capture complexity classes (e.g., \cite{Immerman:1999,Libkin:2004}).
For example, polynomial time queries on ordered relational structures
can be captured
by first order logic augmented with a least fixpoint operator.
Some such results translate easily into programming languages.
\item \emph{Program schemes}
are restricted forms of recursion for which certain properties (e.g.,
termination) are decidable \cite{Courcelle:TCS:1990}.  
\future{
\item Injective or reversible languages (e.g., \cite{Mu:MPC:2004}) capture the property that
no entropy (heat) need be produced while performing a computation, or
equivalently, that the computation is fully reversible.
}
\future{
\item Synchronous languages such as Lucid-Synchrone allow ``correct
by construction'' (implicit capture) of safety properties for
reactive systems.
}
\end{itemize}
We may hope to design metalanguages that 
capture safety properties in a similar way--- this is essentially the
goal of the MetaML research programme \cite{Sheard:unpub:1994,Taha:TCS:2000}.
We can use the tools of computability theory
to reason about when a language capturing a property might exist
and what properties it might have.

In what follows we will consider properties $\psi(g)$ defined by
arithmetical formulas as in the previous section.
\ignore{***
We show a few examples to illustrate that
this is a plausible way to proceed.
\begin{itemize}
\item The property ``$g$ halts on every input'' is defined by the
formula:
\begin{align*}
\psi(g) \leftrightarrow \forall x ~.~ \varphi_g(x) \downarrow
\end{align*}
\item The property ``Every instance is safe'':
\begin{align*}
\psi(g) \leftrightarrow \neg\exists x ~.~ ( \varphi_g(x) \downarrow ) \wedge
(\varphi_g(x) \not\in \mathsf{SafeInstance})
\end{align*}
\item The property ``$g$ halts in $O(n^2)$ steps in the
length of its input'' is defined by the sentence:
\begin{align*}
\psi(g) \leftrightarrow
\exists c ~.~ \exists x_0 ~.~ \forall x \geq x_0 ~.~
\varphi_{g,c \cdot \lceil1+\log_2 x \rceil ^2}(x) \downarrow
\end{align*}
(Here the length of an input $x$ is $\lceil 1+\log_2 x \rceil$, and we
take $g \in O(f)$ to mean there is a constant
$c$ such that eventually $g(n) < f(n)$.)
\end{itemize}
***}
If a particular generator $g$ satisfies a property $\psi(g)$, 
we say ``$\psi(g)$ holds'' or ``$g$ satisfies $\psi$''.

\begin{defn}[Capture]
\label{defn:capture}
We say a restricted metalanguage $L \subset \N$ \emph{captures} a 
property $\psi$ when $L$ is a decidable subset of generators, and
\begin{enumerate}
\item Every program $g \in L$ in the restricted metalanguage
satisfies the property $\psi$:
\begin{align}
\forall g \in L ~.~ \psi(g)
\end{align}
\item For every (unrestricted) generator $g' \in \N$ such that $g'$ satisfies
$\psi$, there is some equivalent (restricted) generator $g \in L$ such 
that $\varphi_g = \varphi_{g'}$:
\begin{align}
\forall g' \in \N ~.~ \psi(g') \rightarrow (\exists g \in L ~.~ \varphi_g = \varphi_{g'})
\end{align}
\end{enumerate}
\end{defn}

For example, if a metalanguage $L$ captures the property ``for all $x$,
$\varphi_g(x)$ runs in $O(|x|^2)$ time'', this means not only that every 
generator in $L$ runs in quadratic time, but also that every computation
that \emph{can} run in quadratic time is expressible in $L$.

First we consider the problem of metalanguages capturing
\emph{functional} properties, such as always generating safe
instances or always terminating.

\begin{defn}[Functional property]
\label{defn:functional}
A property $\psi(g)$ is \emph{functional} when, equivalently:
\begin{enumerate}
\item $\psi$ is
defined solely in terms of termination and input-output behaviour;
\item $\psi$ is an index set;
\item $(\varphi_g = \varphi_{g'}) \rightarrow (\psi(g) \leftrightarrow \psi(g'))$.
\end{enumerate}
\end{defn}
For example, whether a
generator always produces safe instances is a functional property;
whether it runs in quadratic time is not, and is considered a
\emph{non-functional} property.  (The term \emph{non-functional}
is unfortunate but traditional in software engineering.)

One way to design a language capturing a property is to package
programs together with proofs of that property, as in proof-carrying
code \cite{Necula:POPL:1997} or Royer and Case's treatment of
provably bounded programming systems \cite[\S 4.3.1]{Royer:1994}.
This yields a sufficient condition
for the existence of a language capturing a property.
\begin{prop}
\label{prop:proofcap}
Let $\psi$ be a property for which there is a sound 
proof system $\vdash_{\psi}$ with the following properties:
\begin{enumerate}
\item Checking whether a deduction $\mathscr{D}$ is a valid proof in the system
$\vdash_{\psi}$ is decidable;
\item For every generator $g$ satisfying $\psi$, there exists an
equivalent $g'$ such that $\varphi_g = \varphi_{g'}$ and there is 
a deduction in $\vdash_{\psi}$ proving $\psi(g')$.  (This is strictly
weaker than requiring the proof system to be \emph{complete} for $\psi$.)
\end{enumerate}
Then there is a language capturing $\psi$.
\end{prop}
\begin{proof}
We follow the proof-carrying code idea,
designing a language whose every program is a pair $(g,\mathscr{D})$
where $g$ is the generator and $\mathscr{D}$ is a nonexecutable
payload containing a suitably encoded proof of $\psi(g)$ in the system
$\vdash_{\psi}$.  We define the language $L$ to be only those
$(g,\mathscr{D})$ where $\mathscr{D}$ is a valid deduction proving
$\psi(g)$; this is a $\Delta^0_1$ subset because of the premise (1).

We claim this language captures $\psi$ in the sense of \refdefn{defn:capture}:
Every generator in $L$ clearly has the property $\psi$, due to 
soundness of the proof system; and
every generator $g \in \N$ such that $\psi(g)$ holds has an equivalent
program in $L$ by the premise (2).
\qed\end{proof}

\future{If we have a proof that a language $L$ captures $\psi$, the
converse is also true; build $\vdash_\psi$ from the decision procedure
for $L$.}



Languages that require programmers to attach proofs suffer from 
so-called ``technology adoption issues.''
Better perhaps to find a language that \emph{implicitly}
captures the propery; such languages do not require programmers to
explicitly write proofs.  However, writing programs in languages that
capture functional properties does have an \emph{implicit} relationship
to proofs: reworking a generator from a general-purpose language to a
restricted language is tantamount to \emph{proving} the property.

\begin{thm}[Capture is tantamount to proof]
\label{thm:tantamount}
Let $\psi$ be a \emph{functional} property of generators, and $L \subset \N$
a restricted language capturing $\psi$.
Given a generator
$g' \in \N$ satisfying $\psi$, the problem of
transforming it into an equivalent generator $g \in L$ in the restricted
language by means of (provably) semantics-preserving steps
is at least as hard as finding a proof of $\psi(g')$.
\end{thm}
\begin{proof}
We assume we have already a proof that $L$ captures $\psi$.
At each step of transforming $g'$ into $g$ we can,
by means of semantics-preserving steps,
maintain a proof that the two versions are equivalent, so at the
end of the process we have a proof that
$\varphi_g = \varphi_{g'}$.
Since $L$ is decidable we can readily obtain a proof
that $g \in L$ at the end of the process. We then have proofs of
\begin{enumerate}
\item $g \in L$ ($g$ is in the restricted language)
\item $\forall g \in L ~.~ \psi(g)$ (from capture of $\psi$ by $L$)
\item $\varphi_g = \varphi_{g'}$ (from semantics-preserving steps)
\end{enumerate}
From (1) and (2) we obtain $\psi(g)$; from (3) and the premise that
$\psi$ is a functional property we obtain $\psi(g')$.

Therefore the problem of proving $\psi(g')$ is reducibile to
transforming $g'$ into the restricted language $L$ by means of
semantics-preserving steps.\qed
\end{proof}

\future{Maybe Bennett's notion of information distance could be used
here, i.e., when we rework a program into a restricted language it
is inherently more informative, and we have some kind of isometry
between different representations.}

\future{Can we use some kind of argument about the transformation
process not cycling to show that the program has to get really
big?  (But what if the program gets super small in the last step?)
}

\noindent
Related results on succinctness (\refsec{s:succinct}) suggest that
the reworked program may be as long as a proof of the property $\psi(g)$.

\future{Try using incompressibility to show that any rewriting in
the form of \refthm{thm:tantamount} must necessarily make programs
longer than any computable function.}

\begin{cor}
If $\psi$ is an undecidable property, there can be no automated (computable)
process for reworking generators into the restricted language.
\end{cor}

The practical implications of this are that writing generators in
certain restricted languages is just as hard as proving safety properties,
and may require arbitrary creativity. However, there may be an important
social difference: proof construction can be intimidating for programmers,
whereas programming in a restricted language can be a source of interesting
puzzles requiring ingenious solution.
There are intermediate solutions between proof-carrying code and
implicit capture, where we design a language with some mix of explicit
proof and implicit capture of the property.  Type systems are a prime
example: the programmer annotates a program with enough type
information to make type safety easily provable.  From a theoretical 
perspective programmers are constructing proofs of type safety
relative to the decision procedure for the type system in the compiler; 
but it feels more intuitive than formal proof calculi.

\subsection{When is capture possible?}

So we may sometimes find languages that capture undecidable
properties.  In this section we give results on when such languages
may or may not exist.  
There are some niches that can be carved out, though, for example
the sufficient conditions of \refprop{prop:proofcap}.
We also know there are languages capturing
any deterministic time and space bounds due to the existence of
``clocked'' programming systems where programs are annotated with
resource bounds \cite{Royer:1994}.  

Here is a negative result.
We show that arbitrarily hard functional properties cannot
be captured by programming languages, by turning the tables and
characterizing a property in terms of the language capturing it.
\begin{prop}
If there is a language capturing a
\emph{functional} property $\psi$, then $\psi \in \Sigma^0_3$.
\end{prop}
\begin{proof}
From the definition of capture (\refdefn{defn:capture}) and the
key fact that $\psi$ is functional (\refdefn{defn:functional}), we have
the following correspondence.
\begin{align}
\psi(g) \leftrightarrow \exists g' ~.~ \underbrace{(\varphi_g = \varphi_{g'})}_{\Pi^0_2} \wedge \underbrace{(g' \in L)}_{\Delta^0_1}
\end{align}
The relation $\varphi_g = \varphi_{g'}$ is $\Pi^0_2$, so $\psi$ is
$\Sigma^0_3$.
\qed\end{proof}

This implies we cannot capture in languages any functional property
not definable in $\Sigma^0_3$, i.e., the whole span of the arithmetical
hierarchy above $\Sigma^0_3$ is off-limits.  However, this does not rule out
the possibility of arbitrarily hard \emph{nonfunctional} requirements being 
captured.

\ignore{***
It would be handy if we could prove the other direction and say
``If $\phi$ is a $\Sigma^0_3$ functional property, there is a language
capturing it.''  Unfortunately this is not the case.  A counterexample
is given by the existence of so-called \emph{immune} sets in $\Pi^0_1$.

\begin{defn}
An \emph{immune} set is an infinite set containing no infinite
$\Sigma^0_1$ subsets.
\end{defn}

Informally, a set $A$ is immune if any program that prints
out only elements of $A$ can only print out a finite number of elements.
Immune sets are infinite, but seem finite to our
Turing-limited micro-world.

It turns out that complements of `simple' sets are immune, and there
is a simple set in $\Sigma^0_1$, which places an immune set in
$\Pi^0_1$ (\cite{Cooper:2003}).  

\begin{prop}
There is a property $\psi \in \Sigma^0_1$ that cannot be captured
by a programming language.
\end{prop}
\begin{proof}
Let $\Psi$ be an immune set in $\Pi^0_1$.  Then $\Psi$ has infinitely
many elements.  
\end{proof}
***}

Here is a well-known fact that sets the stage for proving we cannot
capture the property ``every generator $g$ always halts and produces 
a safe instance,'' i.e., total correctness of generators, in a language.  

\begin{prop}
\label{prop:notot}
There is no language capturing the total computable functions, i.e., the
property
$\mathrm{Tot}(p) \leftrightarrow \forall x ~.~ \varphi_p(x) \downarrow$.
\end{prop}
\begin{proof}
Suppose $L \subset \N$ is a language
capturing $\mathrm{Tot}$; then every total computable function
is expressible in $L$.
We use diagonalization to construct a total computable function
obviously not in $L$.
Consider an enumeration $\{ p_i \}_{i \in \N}$ of the programs
in $L$; such an enumeration exists since $L$ is decidable by
\refdefn{defn:capture}.
Consider the function
\begin{align*}
f(n) &= 1 + \varphi_{p_n}(n)
\end{align*}
Since $p_n \in L$, this program halts on every input.  It is also
total and computable, and is therefore expressible in $L$.  
Let $k$ be the $L$-program that computes it.  Then 
$\varphi_k(k) = 1 + \varphi_k(k)$,
a contradiction since $z = 1 + z$ has no solution in the
naturals.
\qed\end{proof}

\begin{prop}
If there are at least two safe instances,
there is no metalanguage capturing the property ``$g$ halts on every
input and outputs a safe instance.''
\end{prop}
\begin{proof}
Following the same style of diagonalization argument in \refprop{prop:notot},
changing every safe instance on the diagonal.  (The diagonalization would
fail if there was only one safe instance).
\end{proof}

\ignore{***
Intuitively, we can say there exists a language for $\psi$ whenever
there is a decidable set $L$ giving a system of representatives for 
the equivalence classes $\psi / \sim$, where $\sim$ is program 
equivalence, i.e., $(g \sim g') \leftrightarrow (\varphi_g = \varphi_{g'})$.
In computability theory, an \emph{immune} set is an infinite set 
containing no infinite $\Sigma^0_1$ subsets.
***}

\ignore{***
\begin{thm}
If $\psi / \sim$ is infinite and $\psi$ is immune, there is no 
language capturing $\psi$.
\end{thm}
\begin{proof}
Immunity of $\psi$ implies its intersection with any $\Sigma^0_1$
is finite.  Any nontrivial property has an infinite number of
programs satisfying it.
\end{proof}
***}

\ignore{***
\begin{prop}
Any $\Delta^0_2$ property has a $\Sigma^0_1$ language capturing it.
\end{prop}
\begin{proof}
Use a $\Delta^0_2$ approximating sequence.
\end{proof}
***}

\subsection{Succinctness}

\label{s:succinct}
It is a well-studied phenomena that often when we move from one language to a
more restricted version, some programs have to get larger --- a loss
of \emph{succinctness}.  
The intuitive reasons for this are demonstrated by revisiting the 
language seen earlier
capturing the property ``halts on only finitely many inputs''
by programs of the form:
\begin{align*}
f(x) &= 
\begin{cases}
c_1 & \text{when } x = x_1 \\
c_2 & \text{when } x = x_2 \\
~~\vdots    & ~~~~\vdots  \\
c_n & \text{when } x = x_n \\
\uparrow & \text{otherwise}
\end{cases}
\end{align*}
Consider a program written in a general-purpose language that, given
input $x < 10^{100}$, outputs $x+1$, 
otherwise diverges.  This halts only on finitely many inputs,
and can be implemented in a few lines of C code.
The corresponding ``lookup table'' program in the above language
is too long to fit into the observable universe.  (If we scale
down the exponent we can get the more practical ``too long to
fit into any existing computer.'')  An underlying cause is 
that we can pose problems 
easily solveable in an unrestricted language, but ``look random''
to the restricted language and one cannot do any better than
decomposing it into a large number of cases, as in the above example.
Problems that can be decomposed only into an infinite number of cases
are inexpressible.
\ignore{***
Note also that one can change the upper bound from $10^{100}$ to
any computable quantity, so that the loss in succinctness in going
from a general-purpose language to the above restricted language
cannot be bounded by any computable function of the program length; 
far from being unusual, this turns out to be typical of ``succinctness 
gaps'' between interesting languages.
***}

Let us write $|\cdot|$ for the length of a program; such measures are
usually required to satisfy the very weak axioms of Blum \cite{Blum:IC:1967}.  
Counting bits of the representation is satisfactory.

\begin{defn}[Computably succint]
Suppose $L$ and $L'$ are two languages.  We say $L$ is 
\emph{computably succinct} relative to $L'$ if for
every program $p \in L$ for which there is a functionally
equivalent program in $L'$, there is a $p' \in L'$ such that
\begin{align*}
|p'| \leq f(|p|)
\end{align*}
where $f$ is some computable function.
\end{defn}

Saying one language is not computably succinct relative to another
is a strong statement; for example, it implies that the loss
in succinctness cannot be bounded by your favourite fast-growing
computable function, for example the `power tower':
\begin{align*}
T(k) = \underbrace{10^{10^{10^{10^{\ldots^{10}}}}}}_{k}
\end{align*}
\noindent
where $T(1)=10$, $T(2) = 100$, $T(3) = 10^{100}$, and so on.
(Cosmologists suppose the number of atoms in the observable universe
is less than $T(3)$.)

All the Turing-complete languages are computably succinct relative to
one another, and in fact the interesting ones are all within
an additive constant of one another; this follows from a `two-part
code' construction \cite[\S 2.1]{Li:1997}.

The tradeoff between succinctness and power of languages has been
explored rather exhaustively, and we summarize only some highlights
here.  For details Royer and Case \cite{Royer:1994}
is recommended for the subrecursive languages perspective, 
and Chapter 7 of Li and Vitányi \cite{Li:1997} is recommended for
the Kolmogorov complexity viewpoint (concerned primarily
with \emph{instance complexity} rather than computable functions, 
but still interesting).

Results on succinctness of languages fall loosely into three
classes.  

\begin{enumerate}
\item Loss
in succinctness when moving from one language to more restrictive language.
The general flavour of such results is that if you
restrict a language that can compute at least 
polynomial-time functions in a sufficiently strong way, the resulting 
loss in succinctness cannot
be bounded by any computable function.  The first such result was
achieved by Blum \cite{Blum:IC:1967}, and similar results are
abundant \cite{Royer:1994}.  
\ignore{***There is a powerful
generalization due to Hartmanis \cite{Hartmanis:TCS:1983} that
connects language succinctness to ``G{\"o}del speed-up'' ---
strengthening a proof system to make
some proofs incomputably shorter.***}

\item Losses in succinctness between
two languages of the same expressive power.  For example,
most introductory theory classes
cover the fact that nondeterministic finite automata (NFAs) can be
converted to DFAs with at most an exponential expansion in size.
Note though, that both capture the regular languages.

\item Losses in succinctness when moving back and forth
between two languages of the same expressive power.  For
instance Hartmanis gives an example
of two different languages capturing PTIME, neither of which
is computably succinct relative to the other \cite{Hartmanis:TCS:1983}.
\end{enumerate}

\ignore{***
\editor{The NFA/DFA example is illustrative because it shows that
there can be a change in succinctness between two formalisms of the
same expressive power.  This is akin to introducing new axiom schemes in
a proof system whose instantiations are provable in that system, but
at great cost.  The other kind of result is on succinctness results
between two formalisms, e.g., second-order vs. first-order logic.
When we get up to Turing-complete languages we suddenly have the
situation where no axiom schemes offer any benefit in succinctness,
and there are no more powerful formalisms.  Note also that the
NFA-DFA explosion happens a.e. under uniform measure, think of
NFA as a relational structure and almost surely there exist
exponential-blowing-up substructures?}
***}

\ignore{***
The following is a variant of a theorem due to
Hartmanis \cite{Hartmanis:TCS:1983} giving a necessary
condition for a restricted language to be succinct.

\editor{Hartmanis's proof is only for total languages.}
\begin{thm}[Hartmanis]
Let $L \subset \N$ be a restricted programming language,
and $\Delta = \N - \indexclosure{L}$ the programs
inexpressible in $L$.
If $\Delta \not\in \Sigma^0_1$, then $L$ is not computably succinct.
\end{thm}
\begin{proof}
(By contradiction.)
Suppose such an $f$ existed.  Then
\begin{align*}
\N - \indexclosure{L} &= \{ p ~|~ \exists p' \in L ~.~ \exists x ~.~ \varphi(p) \neq \varphi(p') \}
\end{align*}
\end{proof}
***}


With respect to languages capturing properties, we can make the
following observation.
\begin{thm}
If a language $L \subset \N$ is computably succinct, then any functional
property $\psi$ it captures is $\Pi^0_2$.
\end{thm}
\begin{proof}
Suppose $L$ captures a property $\psi$.  As before we turn the tables
and define $\psi$ in terms of the language:
\begin{align}
\psi(g) \leftrightarrow \exists g' ~.~ (\varphi_g = \varphi_{g'}) \wedge (g' \in L)
\end{align}
Since $L$ is computably succinct, there is some computable $f$ such that
\begin{align}
\psi(g) \leftrightarrow \exists g' \leq f(g) ~.~ \underbrace{(\varphi_g = \varphi_{g'}) \wedge (g' \in L)}_{\Pi^0_2}
\end{align}
The addition of $\leq f(g)$ turns the existential quantifier into a
\emph{bounded} quantifier; therefore $\psi(g) \in \Pi^0_2$.
\qed\end{proof}

\ignore{***
Comment.  The succinctness loss from a universal to a restricted language
does not hold in an almost sure sense, unless we go out of our way to
make the restricted language verbose.  The reasoning is as follows.
Enumerate the valid programs of the restricted language $L$.
We know almost every position in this enumeration is random.
For any program $p$ in the restricted language $L$, we can 
find a generator for it in the universal language of size $\leq C(i) + c$.  
where $i$ is the position of $p$ in the enumeration.  But since $i$
is almost surely incompressible, $C(i) \approx \log_2(i)$.
***}

\section{Chasing properties}
 

When our attempts to capture a property by a language fail,
there remains the possibility of approximating the property.
For example, although total correctness
of generators cannot be captured by a language, we can choose
some restricted language as a starting point and gradually
build it up so as to increase its power and succinctness.

\ignore{***
One often encounters in practice undecidable problems of great importance,
such as designing languages that halt on every input.  Such languages are
restricted in expressiveness, so one sees in the literature a sequence
of every-stronger formalisms that have the property and are every more
complicated to implement.  
***}

We can model this process by a chain of languages
$L_0 \subset L_1 \subset L_2 \cdots$ converging towards
the desired safety property $\psi$.  Note that given
$L_i$ we can always find a set $L_i \subset L_{i+1} \subset \psi$,
for example, by adding a finite number of special cases to the
test for $L_i$.  But in practice we find a slightly stronger
proof system, capture some common patterns, and so forth.
We call this \emph{chasing} a property.

Let us write $C(L_i)$ for the length of the shortest
program deciding the set $L_i$.  

\ignore{***
\begin{prop}
If $L_0 \subset L_1 \subset L_2 \cdots$ is a countable (but not
c.e.) sequence of languages such that
\begin{enumerate}
\item Each language $L_i$ has the property $\psi(p)$ for every $p \in L_i$;
\item $\lim_{i \rightarrow \infty} L_i = \{ p ~|~ \psi(p) \}$, i.e.,
in the limit we recover the language with the property $\psi(p)$;
\end{enumerate}
Then, $\lim_{i \rightarrow \infty} C(L_i) = \infty$, i.e., the length of
program required to decide the language $L_i$ tends to infinity.
\end{prop}
\begin{proof}
By contradiction.  Suppose $\lim_{i \rightarrow \infty} C(L_i) = k$
for some finite $k$.  Then there would be a language capturing $\psi(p)$,
violating \refprop{prop:haltlang}.
\end{proof}
***}

\ignore{***
\noindent
The above proposition tells us that \emph{if} there is a sequence
of approximations to a language capturing $\psi(p)$, they have
to get harder and harder to describe.  However, it does not tell
us that such an approximation exists.  It turns out that if the
property is $\Delta^0_2$ --- that is, only ``slightly'' undecidable ---
they we are guaranteed the existence of an approximating sequence
of languages.  In other words, we can get arbitrarily close
to a language capturing $\psi(p)$, at the price of ever-more-complicated
languages.
***}

\begin{prop}
Let $\psi$ be an undecidable property of generators, and
$L_0 \subset L_1 \subset L_2 \cdots$ a countable (but not c.e.) sequence
of languages, each decidable, such that:
\begin{enumerate}
\item Each language $L_i$ has the property $\psi(p)$ for every $p \in L_i$;
\item $\bigcup_{i \in \N} L_i = \psi$, i.e.,
in the limit we recover exactly the language $\psi$;
\end{enumerate}
Then, $\lim_{i \rightarrow \infty} C(L_i) = \infty$, i.e., the length of
program required to describe the language $L_i$ diverges.
\end{prop}

\begin{proof}
If $C(L_i)$ did not diverge, we would have a finite program deciding
the undecidable property $\psi$, a contradiction.
\end{proof}

So, in essence the best one can hope for is to capture the perfect
metalanguage in the limit by approximating it from below with 
ever-more-complicated languages.  This is the conservative
approach.

\future{When is it possible to approximate a language in a limit
in this sense?  Relation to approximation sequences from computability
theory.}

An optimistic approach is to start with the universal language
$L_0 = \N$, which fails the safety property, and find languages
$L_0 \supset L_1 \supset L_2 \supset \cdots$ 
that gradually winnow out the unsafe cases and converge
toward $\psi$, with ever-more-complicated languages.

These two approaches --- approximating $\psi$ conservatively from below
or optimistically from above --- appear to represent irreconcilable 
approaches to metaprogramming language design with no middle ground.
I would propose Meta\-ML as representative of the first, and C++ as
emblematic of the second, particularly the $C(L_i)\rightarrow \infty$ part.

\ifverbose{
\section{When is ``Going Meta'' a jump of Turing degree?}
\label{s:jump}

When we make the shift in perspective from programming
to meta-programming --- ``going meta,'' as it were,
what happens to the difficulty of verification problems?
How does the hardness of these problems relate:
\begin{enumerate}
\item Does a program $p$ satisfy a property $\psi$?
\item Does every program generated by a metaprogram $g$
satisfy a property $\psi$?
\end{enumerate}

These questions suggest a connection to the \emph{jump operator}
in computability theory.  
We define an operator on properties in the following way:
\begin{defn}[Meta-jump]
Let $\psi(p)$ be a property of programs.  The meta-jump of $\psi$
is the property $\psi^\dagger$ given by:
\begin{align*}
\psi^\dagger(g) &\Leftrightarrow \forall y ~.~ \underbrace{(\varphi_g(y) \downarrow)}_{\text{g halts}} ~~ \rightarrow \underbrace{\psi(\varphi_g(y))}_{\text{$\psi$ holds for its output}}
\end{align*}
i.e., we shift from the property of a single program $\psi(p)$ to
the property $\psi(p)$ holding for every program $p$ produced by
the metaprogram $g$.
\end{defn}

We might expect $\psi^\dagger$ to be much harder to decide
than $\psi$, but this appears not to be the case.

\iflong{
Two quick examples demonstrate the issue
is not clear-cut.

\begin{enumerate}
\item Let $\psi$ be the property ``halts on at least one input.''
Then $\psi^\dagger$ is the property ``every program output by the generator
halts on at least one input.''
This is a very desirable property for a generator to have;
otherwise it would only produce programs that went into an
endless loop without producing output.  In symbols,
\begin{align}
\psi(p) &\Leftrightarrow \exists x ~.~ (\varphi_p(x) \downarrow) \\
\psi^\dagger(g) &\Leftrightarrow \forall y ~.~ (\varphi_g(y) \downarrow) \rightarrow \exists x ~.~ \varphi_{\varphi_g(y)}(x) \downarrow
\end{align}
The property $\psi$ is computably
enumerable or $\Sigma^0_1$.  The property $\psi^\dagger$ is
$\Pi^0_2$, and not $\Sigma^0_1$, so
$\psi^\dagger$ might be much harder than $\psi$, as expected.
\item Let $\psi$ be the property ``halts on every input.''  The
property $\psi^\dagger$ is ``every program output by the generator halts
on every input,'' useful if you are writing generators for
safety-critical systems.
\begin{align}
\psi(p) &\Leftrightarrow \forall x ~.~ \varphi_p(x) \downarrow \\
\psi^\dagger(g) &\Leftrightarrow \forall y ~.~ (\varphi_g(y) \downarrow) \rightarrow \forall x ~.~ \varphi_p(x) \downarrow
\end{align}
The property $\psi(p)$ is $\Pi^0_2$; the property $\psi^\dagger(g)$ is also
$\Pi^0_2$.  So in this case the meta-jump does not cause us to move up a
level in the arithmetical hierarchy.
\end{enumerate}
}

Let us use formula syntax to roughly characterize the action of the meta-jump operator.
The meta-jump adds a $\forall$ quantifier at the start of a formula.
Based on this syntactic characterization we can immediately get:
\begin{prop}
The meta-jump operator satisfies the following rules:
\begin{enumerate}
\item If $\psi \in \Sigma^0_k$ then $\psi^\dagger \in \Pi^0_{k+1}$.
\item If $\psi \in \Pi^0_k$ then $\psi^\dagger \in \Pi^0_k$.
\end{enumerate}
\end{prop}
Note that (2) implies that the meta-jump has no effect on the
placement of formulas that are $\Pi^0_k$ but not $\Delta^0_k$.
The condition (1) states that it is possible for $\Sigma^0_k$ properties
to become harder under the meta-jump, but this is not necessary
because of the inclusion
$\Delta^0_k \subset \Sigma^0_k \subset \Pi^0_{k+1}$.
(Note that this characterization is very rough and does not
consider finer degrees of undecidability such as $m$-degrees).

It turns out that the safety properties usually talked about
in connection with metaprogramming and program generation are all
in the form $\Pi^0_k$: halting, always producing a safe instance,
both halting and producing a safe instance, for example.
We usually conceive of a safety property
as preventing certain some (usually infinite) set of
\emph{failure conditions} from occurring.  We can enumerate
the failure conditions and check them one by one; if
a failure condition occurs we know the safety property
fails.  This style of safety condition is always $\Pi^0_1$
relative to some relation, and therefore 
there is no jump in the level of the arithmetical hierarchy when
we `go meta.'  

\ignore{***
However, we can prove the following:
\begin{prop}
If $\psi \in (\Sigma^0_k - \Delta^0_k)$ then
$\psi^\dagger \not\in \Sigma^0_k$.
\end{prop}
***}

\ignore{***
Perhaps counter-intuitively, meta-meta-programming does not cause
any further jumps: $\psi^{\dagger\dagger}$ has the same level in the
arithmetical hierarchy as $\psi^\dagger$.
***}

\ignore{***
Want to express: If $\psi \in \Sigma^0_k$ but not definable in $\Delta^0_k$,
then $\psi'$ is not definable in $\Sigma^0_k$.

Harder version of jump:
\begin{thm}
If $\emptyset' \leq_T \psi$ then $\emptyset'' \leq_T \psi'$.
\end{thm}

Examples of $\Sigma^0_k$ properties: the program halts for at least
one input.

Prove that there is a real jump of turing degree, not just a
jump in syntactic class.

\subsection{Practical Implications}

What does this mean at the lowest level of the hierarchy?  What
properties can we decide of generators?

Give an example of a property that becomes inapproximable when we
``go meta.''

***}

}

\section{Conclusions}

We set out to characterize tradeoffs in metaprogramming, in retrospect
a somewhat presumptuous goal, since the problems turn out to be unexpectedly
deep and requiring further investigation.  The computability approach is
very useful in providing quick and coarse characterizations of tradeoffs.

One result that remains elusive is a characterization of the
tradeoff between safety and expressive
power involved in metalanguages having the property
``every instance is type-safe.''  
The primary challenge is that in a general-purpose
metalanguage one can always satisfy the capture
properties vacuously by introducing layers of interpretation, and a way 
to disallow these `cheats' is not apparent without dropping to a 
subrecursive metalanguage.  Partial correctness
of metaprograms is a `run-time manageable' property, in the
sense that one can detect bad outputs before they happen.

Despite these difficulties a rough picture of the major 
tradeoffs in `safe metaprogramming' emerges.
\begin{enumerate}
\item Total correctness of generators cannot be captured.  If termination
of the generator is required, one must pick a suitable subrecursive
language and try to recoup succinctness and power as needed by building
in provably safe `escapes' back up to more powerful classes of
generators.
\future{It would be interesting to use information-flow security to allow
access to higher-computation oracles that only informed path-choice
decisions, as in complexity theory.  However, then you would have to
prove that the algorithm would behave the same regardless of the oracle
input; that might be hard to prove.
}
\item Whether there are languages capturing partial correctness of
metaprograms in a useful way is uncertain.  However, we can say that if such languages
exist, and are not vacuous, then we expect because of \refthm{thm:tantamount}
that reworking generators into such languages is as hard as \emph{proving}
partial correctness.  However for social reasons `capture' is more
desirable, if it can be achieved.
\item When we restrict the power of metalanguages, we lose the
ability to compress certain kinds of motifs and patterns.  Whether
this causes a practical (rather than theoretical) loss in succinctness
depends on the characteristics of the problem domain (cf. \cite{Veldhuizen:LCSD:2005}).
\end{enumerate}

To summarize, we can say that there is choice between two strategies
when designing a metaprogramming language.  One can give safety
primacy, start with a restricted language, and try to build up
the complexity of the language so as to recapture lost power and
succinctness.  Or, one can give power primacy, start with a universal
language, and try to build up the complexity of the language so as
to capture necessary safety properties.  These represent fundamentally
different attitudes toward metaprogramming.

\section*{Acknowledgments}

I am grateful to Kyle Ross and Jeremiah Willcock for suggesting improvements,
and Saleh Aliyari for his expert help with computability theory.

{\footnotesize
\bibliographystyle{abbrv}
\bibliography{KEYWORDS,abint,activelib,alg,algebra,alias,arrays,asymptotics,autocomplexity,automata,checking,coalgebra,cocv,coding,combinatorics,compilers,complexity,components,computability,cpp,dsl,extensible,fixpoint,generics,graph,hardware,hashing,hci,historyofcomputing,incremental,java,kolmogorov,lang,lattice,libraries,logic,matrixmult,meta,nonstandard,numanalysis,numtheory,oon,order,ordinals,parallel,parsing,partial,persistent,philosophy,prob,processes,realtime,recurrence,representations,research,reuse,rewrite,safety,security,selection,semantics,settheory,slicing,software,solvers,staging,stats,steiner,stores,subrecursive,tcs,theories,tveldhui,types,verify,writing,zipf,reversible}

\begin{thebibliography}{10}

\bibitem{Blackwell:CT:2001}
A.~F. Blackwell, C.~Britton, A.~Cox, T.~Green, C.~Gurr, G.~Kadoda, M.~S. Kutar,
  M.~Loomes, C.~L. Nehaniv, M.~Petre, C.~Roast, C.~Roes, A.~Wong, and R.~M.
  Young.
\newblock Cognitive dimensions of notations: Design tools for cognitive
  technology.
\newblock {\em Lecture Notes in Computer Science}, 2117:325--341, 2001.

\bibitem{Blum:IC:1967}
M.~Blum.
\newblock On the size of machines.
\newblock {\em Information and Control}, 11(3):257--265, Sept. 1967.

\bibitem{Brosgol:SIGPLAN:1971}
B.~M. Brosgol.
\newblock An implementation of {ECL} data types.
\newblock {\em SIGPLAN Not.}, 6(12):87--95, 1971.

\bibitem{Ceruzzi:1998}
P.~E. Ceruzzi.
\newblock {\em A History of Modern Computing}.
\newblock MIT Press, Cambridge, Mass., 1998.

\bibitem{Charikar:STOC:2002}
M.~Charikar, E.~Lehman, D.~Liu, R.~Panigrahy, M.~Prabhakaran, A.~Rasala,
  A.~Sahai, and abhi shelat.
\newblock Approximating the smallest grammar: Kolmogorov complexity in natural
  models.
\newblock In {\em STOC '02: Proceedings of the thiry-fourth annual ACM
  symposium on Theory of computing}, pages 792--801, New York, NY, USA, 2002.
  ACM Press.

\bibitem{Cooper:2003}
B.~S. Cooper.
\newblock {\em Computability Theory}.
\newblock Chapman \& Hall/CRC mathematics, 2003.

\bibitem{Courcelle:TCS:1990}
B.~Courcelle.
\newblock Recursive applicative program schemes.
\newblock In {\em Handbook of Theoretical Computer Science, Volume B: Formal
  Models and Sematics (B)}, pages 459--492. 1990.

\bibitem{Czarnecki:2000}
K.~Czarnecki and U.~W. Eisenecker.
\newblock {\em {Generative Programming: Methods, Tools, and Applications}}.
\newblock Addison-Wesley, 2000.

\bibitem{Glueck:PLILP:1995}
R.~Gl{\"u}ck and J.~J{\o}rgensen.
\newblock Efficient multi-level generating extensions for program
  specialization.
\newblock In S.~D. Swierstra and M.~Hermenegildo, editors, {\em Programming
  Languages: Implementations, Logics and Programs (PLILP'95)}, volume 982 of
  {\em Lecture Notes in Computer Science}, pages 259--278. Springer-Verlag,
  1995.

\bibitem{Green:HCI:1989}
T.~R.~G. Green.
\newblock Cognitive dimensions of notations.
\newblock In {\em Proceedings of the HCI'89 Conference on People and Computers
  V}, Cognitive Ergonomics, pages 443--460, 1989.

\bibitem{Hartmanis:TCS:1983}
J.~Hartmanis.
\newblock On {G{\"o}del} speed-up and succinctness of language representations.
\newblock {\em Theoretical Computer Science}, 26:335--342, 1983.

\bibitem{Holloway:SIGPLAN:1971}
G.~H. Holloway.
\newblock Interpreter/compiler integration in {ECL}.
\newblock {\em SIGPLAN Not.}, 6(12):129--134, 1971.

\bibitem{Hopcroft:TMC:1971}
J.~E. Hopcroft.
\newblock An {$n \log n$} algorithm for minimizing the states in a
  finite-automaton.
\newblock In Z.~Kohavi, editor, {\em Theory of Machines and Computations},
  pages 189--196. Academic Press, 1971.

\bibitem{Huang:GPCE:2005}
S.~S. Huang, D.~Zook, and Y.~Smaragdakis.
\newblock Statically safe program generation with {SafeGen}.
\newblock In {\em Generative Programming and Component Engineering ({GPCE})},
  Sept. 2005.

\bibitem{Immerman:1999}
N.~Immerman.
\newblock {\em Descriptive complexity}.
\newblock Graduate Texts in Computer Science. Springer-Verlag, New York, 1999.

\bibitem{Jones:1997}
N.~D. Jones.
\newblock {\em Computability and Complexity from a Programming Perspective}.
\newblock Foundations of Computing. MIT Press, Boston, London, 1 edition, 1997.

\bibitem{Jones:TCS:1999}
N.~D. Jones.
\newblock {LOGSPACE} and {PTIME} characterized by programming languages.
\newblock {\em Theoretical Computer Science}, 228:151--174, 1999.

\bibitem{Li:1997}
M.~Li and P.~Vitányi.
\newblock {\em An introduction to Kolmogorov complexity and its applications}.
\newblock Springer-Verlag, New York, 2nd edition, 1997.

\bibitem{Libkin:2004}
L.~Libkin.
\newblock {\em Elements of finite model theory}.
\newblock Texts in Theoretical Computer Science. An EATCS Series.
  Springer-Verlag, Berlin, 2004.

\bibitem{Liskov:SIGPLAN:1974}
B.~Liskov and S.~Zilles.
\newblock Programming with abstract data types.
\newblock {\em SIGPLAN Not.}, 9(4):50--59, 1974.

\bibitem{Meyer:ACM:1967}
A.~R. Meyer and D.~M. Ritchie.
\newblock The complexity of loop programs.
\newblock In {\em Proceedings of the 1967 22nd national conference of the ACM},
  pages 465--469. ACM Press, 1967.

\bibitem{Necula:POPL:1997}
G.~C. Necula.
\newblock Proof-carrying code.
\newblock In {\em POPL '97: Proceedings of the 24th ACM SIGPLAN-SIGACT
  symposium on Principles of programming languages}, pages 106--119, New York,
  NY, USA, 1997. ACM Press.

\bibitem{Rice:TAMS:1953}
H.~G. Rice.
\newblock Classes of recursively enumerable sets and their decision problems.
\newblock {\em Trans. Amer. Math. Soc.}, 74:358--366, 1953.

\bibitem{Royer:1994}
J.~S. Royer and J.~Case.
\newblock {\em Subrecursive Programming Systems: Complexity and Succinctness.}
\newblock Birkhauser, 1994.

\bibitem{Sheard:unpub:1994}
T.~Sheard and J.~Hook.
\newblock Type safe meta-programming.
\newblock Unpublished manuscript, Oregon Graduate Institute, November 1994.

\bibitem{Siegal:ACM:1959}
H.~Siegal and J.~Painter.
\newblock The use of generators in {TAC}.
\newblock In {\em ACM '59: Preprints of papers presented at the 14th national
  meeting of the Association for Computing Machinery}, pages 1--4. ACM Press,
  1959.

\bibitem{Soare:BSL:1996}
R.~I. Soare.
\newblock Computability and recursion.
\newblock {\em The Bulletin of Symbolic Logic}, 2(3):284--321, Sept. 1996.

\bibitem{Taha:TCS:2000}
W.~Taha and T.~Sheard.
\newblock {MetaML} and multi-stage programming with explicit annotations.
\newblock {\em Theoretical Computer Science}, 248(1--2):211--242, Oct. 2000.

\bibitem{Veldhuizen:LCSD:2005}
T.~L. Veldhuizen.
\newblock Software libraries and their reuse: Entropy, {Kolmogorov} complexity,
  and {Zipf's} law.
\newblock In {\em OOPSLA 2005 Workshop on Library-Centric Software Design
  (LCSD'05)}, 2005.
\newblock arXiv:cs.SE/0508023.

\bibitem{Wegbreit:SIGPLAN:1971}
B.~Wegbreit.
\newblock An overview of the {ECL} programming system.
\newblock {\em SIGPLAN Not.}, 6(12):26--28, 1971.

\end{thebibliography}

}
\end{document}